\documentclass[apj]{emulateapj}
\newcommand{\xmm}{{\em XMM-Newton}}

\shorttitle{X-ray emission from HESS J1731-347/SNR G353.6-0.7}
\shortauthors{Tian et al.}

\begin{document}

\title{X-ray emission from HESS J1731-347/SNR G353.6-0.7 and Central Compact Source XMMS J173203-344518}
\author{W.W. Tian\altaffilmark{1,2}, Z. Li\altaffilmark{3}, D.A. Leahy\altaffilmark{1}, J. Yang\altaffilmark{4}, X.J. Yang\altaffilmark{5}, R. Yamazaki\altaffilmark{6}, D. Lu\altaffilmark{4}} 
\altaffiltext{1}{Department of Physics \& Astronomy, University of Calgary, Calgary, Alberta T2N 1N4, Canada, wtian@ucalgary.ca}
\altaffiltext{2}{National Astronomical Observatories, CAS, Beijing 100012, China; tww@bao.ac.cn}
\altaffiltext{3}{Harvard-Smithsonian Center for Astrophysics, 60
  Garden Street, Cambridge, MA 02138, zyli@head.cfa.harvard.edu}
\altaffiltext{4}{Purple Mountain Observatory, CAS, Nanjing 210008, China}
\altaffiltext{5}{Department of Physics, University of XiangTan, China}
\altaffiltext{6}{Department of Physical Science, Hiroshima University, Higashi-Hiroshima 739-8526, Japan}
\begin{abstract}
We present new results of the HESS J1731-347/SNR G353.6-0.7 system
from {\xmm} and Suzaku X-ray observations, and Delinha CO
observations. We discover extended hard X-rays coincident with the bright,
extended TeV source HESS J1731-347 and the shell of the radio SNR.
We find that spatially-resolved X-ray spectra can generally be
characterized by an absorbed power-law model, with photon-index of
$\sim$2, typical of non-thermal emission.    
A bright X-ray compact source, XMMS J173203-344518, is also detected
near the center of the SNR.
We find no evidence of a radio
counterpart or an extended X-ray morphology for this source, making it
unlikely to be a pulsar wind nebular (PWN).  The spectrum of the
source can be well fitted by an absorbed blackbody with a temperature
of $\sim$0.5 keV plus a power-law tail with a photon-index of
$\sim$5, reminiscent of the X-ray emission of a magnetar. 
CO observations toward the inner part of the HESS source reveal a
bright cloud component at -20$\pm$4 km s$^{-1}$, which is likely located at the same 
distance of $\sim$ 3.2 kpc as the SNR. Based on the probable association between the X-ray and $\gamma$-ray emissions and likely association between the CO cloud and the SNR, we argue that the extended TeV emission originates from
the interaction between the SNR shock and the adjacent CO clouds rather than from a PWN.     
\end{abstract}


\section{Introduction}
The concept that astrophyscal shock waves are efficient accelerators of cosmic rays have widely been accepted for a long time \citep*{bla87, mal01}. Due to the supernova energetics and rates, it seems reasonable that most of Galactic cosmic rays up to 10$^{15}$ eV originate from supernova remnants (SNRs) shocks accelerating particles (protons and electrons) in the interstellar (ISM) and circumstellar medium (CSM). An efficiency of order 10$\%$ of the mechanical energy in the shocks must go into accelerating protons and nuclei to explain the observed intensity of cosmic rays.  However, the relative efficiency of acceleration of protons vs. electrons as well as the maximum energies obtained are not well understood, but are all fundamental to the study of cosmic rays and their effects on the ISM.

Recently X-ray and $\gamma$-ray observations to SNRs, e.g. SN1006, RX J1713.7-3946 and Vela Junior (RX J0852.0-4622 etc) have revealed that SNR shocks are able to accelerate particles to TeV energies \citep{emo02, aha06}. The high energy particles responsible for emitting TeV photons can be either hadrons or leptons, which are produced in astrophysical accelerators as in SNRs. Two primary TeV photon emission processes are the decay of neutral pions produced by interactions of hadronic particles (mostly protons) with ambient matter and the inverse-Compton scattering of electrons on ambient photons (mostly the micro-wave background radiation in the typical ISM).  The particles, which undergo such energy losses,  do not escape the acceleration and emission sites due to the presence of magnetic fields.  However, the TeV photons escape directly, allowing us to explore the sites of these processes.

The High Energy Stereoscopic System (HESS), with its excellent
sensitivity and spatial resolution in the standard of $\gamma$-ray
astronomy, greatly stimulates studies of very high energy astrophysics
in recent years. A multi-wavelength approach 
has proven to be robust in shedding light on the nature of TeV sources. 
Among well-identified radio/X-ray counterparts of about 30 Galactic TeV sources{\footnote{http://tevcat.uchicago.edu/, http://www.mppmu.mpg.de/~rwagner/sources/}},
young shell-type SNRs and pulsar wind nebulae are thought to be two major
$\gamma$-ray generators \citep{bam03,kat05,aha06,uch07,cha08,cam09}. Other candidates have also been proposed as counterparts of some unidentified TeV sources, e.g., old SNRs \citep{yam06,fan08}, hypernova and $\gamma$-ray burst remnants \citep*{iok09}, giant molecular clouds \citep{but08,bam09}, and young stellar clusters \citep{aha07}.
Recently, the extended TeV source HESS J1731-347 is found to almost entirely
overlap an old radio/X-ray SNR candidate G353.6-0.7 \citep[hereafter T08]{aha08, tia08}. The association between
HESS J1731-347
and G353.6-0.7, as suggested by T08, makes this system 
a favorable laboratory for studying the generation of $\gamma$-rays in
evolved SNRs. T08 considered the case of hadronic
particle acceleration in an SNR shock encountering a dense molecular cloud,
based on the theoretical model \citep{yam06}, but
conclusion remained to be drawn due to limitation 
 of the data explored in their paper, i.e., low counting statistics
of the {\sl ROSAT} observation and the lack of high-resolution $^{12}$CO
maps. In this paper, we report new results
obtained from {\xmm} and {\sl Suzaku} X-ray observations and CO
spectral-line data from Delinha 13.7 m radio telescope, and examine the theoretical model. 

\section{X-ray and Radio Observations}
\subsection{{\xmm} and Suzaku Data}
HESS J1731-347 was observed by {\xmm} on March 21, 2007
(ObsID 0405680201; PI: G. Puehlhofer), with a 25 ksec exposure.
We reduced the data obtained from the European Photon Imaging Camera
(EPIC), using the {\xmm}
Science Analysis System (SAS), version 8.0. We selected EPIC-MOS and
EPIC-pn events with pattens 0-12 and 0-4, respectively.
An examination of the light curve indicated that the observation was
contaminated by background flares. Cleaning of the high particle
background
results in effective exposure of 15.9, 11.8 and 6.2 ks for the MOS1,
MOS2 and pn detectors, respectively.
Moreover, the observation was also contaminated by straylight{\footnote{http://xmm.esac.esa.int/external/xmm\_user\_support /documentation/uhb/node23.html}},
presumably from a bright source, 1RXS J173157.7-335007 \citep{vog99}, located at $\sim$50$^\prime$ off axis to the north and outside
the field of view (FoV). Consequently, a substantial portion of
the upper (northern) FoV of the MOS1 and pn detectors was
contaminated (contamination in the MOS2 FoV is apparently minimal) 
and thus masked out from subsequent analysis.
We produced counts and exposure maps for the three detectors, in the
0.8-1.5, 1.5-2.2 and 2.2-7 keV bands. The low energy cutoff is
justified by the relatively high foreground absorption near the
Galactic plane.
We then merged the counts and exposure maps, accounting for the
difference in the effective area among the three detectors.
We also produced corresponding background maps, using the Filter Wheel
Closed data that characterizes the quiescent particle
background.

HESS J1731-347 was observed by Suzaku on Feb. 23, 2007 (ObsID
401099010; PI: G. Puehlhofer), with both X-ray Imaging Spectrometers (XIS) \citep{koy07} and Hard X-ray
Detector \citep{tak07}. The net exposure time is about 33 ks. In this work, we focus on the XIS Front-Illuminated (FI) CCDs that have very high
efficiency and low background especially around 5 keV. Since the XIS2 became
inoperable since November 2002, only data from XIS0 is used here.  We used cleaned version 2.0 data (reduction by HEADAS software version 6.5).  
The Suzaku FoV is 18$\times$18 arcmin$^{2}$, centered at a bright compact source [17$^{\rm h}$32$^{\rm m}$10$^{\rm s}$, -34$^\circ$46$^{\prime}$]. The Suzaku spectra were extracted from a circular-source region with off-source annulus background region in the same observation so that possible contamination by SNR emission is reduced to the minimum. In order to increase the statistics, the spectra from XIS0 and XIS3 were jointly fitted for subsequent analysis.

\subsection {CO Data}
We observe the G353.6-0.7/HESS J1731-347 system by employing the 13.7 meter Delinha millimeter telescope at the Purple Mountain Observatory in March 2008. The telescope has an angular resolution of 55$''$ at the observing frequencies. Simultaneously the three J=1-0 CO isotopic lines (i.e., $^{12}$CO, $^{13}$CO, and $^{18}$CO) were observed by using a cryogenic superconducting SIS receiver and
a multi-line backend spectroscopic system \citep{zuo04}, but only the $^{12}$CO was bright enough for significant detection.  The system provides a velocity coverage from V$_{LSR}$= -150 to 60 km s$^{-1}$ and velocity resolution of 0.37 km$^{-1}$ in $^{12}$CO. Three target spots (see Fig.1b), each with a size of 5$'\times5'$, were selected over the G353.6-0.7 area, which includes the center position at the TeV $\gamma$-ray peak (S1:[RA=17:32:00, Dec=-34:42:00]), and the
{\sl ROSAT} X-ray peak sites (S2: [173220, -345400], S3:[173250, -344600]), respectively. These regions were mapped with a grid size of $1'\times1'$. Each point was integrated up to 6 minutes, resulting an average rms noise level in the spectra to be 0.2-0.3~K. Spectral line data were processed by the CLASS package of GILDAS software developed by IRAM.

\section{Results}
\subsection{X-ray emission}
\subsubsection{X-ray morphology}
Fig.~1 shows the overall X-ray morphology 
revealed by {\xmm}, along with the radio
continuum emission of G353.6-0.7 (Fig.~1a) and the $\gamma$-ray emission 
of HESS J1731-347 (Fig.~1b). 
The radio SNR has a shell-like morphology and an extent of
$\sim$30$^\prime$ in diameter, largely overlapping the extended HESS source.
On the eastern half of the SNR, X-ray emission coincident with the
radio emission was detected in an early {\sl ROSAT} observation
(T08). This is confirmed by the present {\xmm} observation,
although the lower (southern) part of the radio shell falls outside
the {\xmm} FoV. It is also evident that X-ray emission
is present along the western half of the radio shell, which was not detected
in the {\sl ROSAT} observation albeit its larger FoV. 
This can be understood, as the emission is detected only in the
2.2-7 keV band that is almost entirely beyond the {\sl ROSAT} energy coverage.

A number of X-ray substructures are revealed under the moderate
spatial resolution of {\xmm}.
In particular, a prominent filament is present within the shell,
passing through the brightest (inner) part of the HESS source 
(Fig.~1b) and partially coincident with radio continuum emission (Fig.~1a).
With comparable widths of $\sim$$2^\prime$, the filament and the
eastern X-ray shell together define a ring-like feature 
(enclosed by the two dotted circles in Fig.~1a; hereafter
referred to as the ring), along which there are several bright knots 
(highlighted by the small ellipses in Fig.~1a).
Several plumes are present northwest of the ring, where the radio shell
apparently breaks. From their projected positions, it is not clear whether 
the plumes are the natural extent of the shell or the filament.
Furthermore, there is a bright compact source centering at (R.A.,
Dec.) = (17$^{\rm h}$32$^{\rm m}$03$^{\rm s}$, -34$^\circ$45$^\prime$18$^{\prime\prime}$), showing no counterpart on the radio image. This source is named XMMS
J173203-344518 hereafter.    

The ring is not necessarily a coherent feature. Nevertheless, subsequent
analysis is focused on the X-ray emission along the ring, as it
occupies the central portion 
of the {\xmm} FoV that is not subject to the straylight contamination.
Fig.~2 shows the azimuthal distribution of the X-ray
emission along the ring. Individual peaks, e.g., at position angles 
of $\sim$120$^\circ$, 190$^\circ$ and 230$^\circ$, arise from the
bright knots (hereafter referred to as K1, K2 and K3). 
The ring appears harder on its western half (i.e., the filament, in angular
range of 0$^\circ$-180$^\circ$) than on its eastern half (i.e., the shell).

\subsubsection{X-ray spectra}
We will focus on the {\xmm} data because Suzaku XIS has a smaller FoV and a lower spatial resolution ($\sim$ 1.8 arcmin) than that of the {\xmm}.
We extract spectra, based on the MOS1, MOS2 and pn data, from various regions along the ring, including the three knots, the eastern half of the ring (K2 and K3 excluded), and the western half of the ring
(i.e., the filament, K1 excluded). As the SNR fills almost the entire
FoV, an ideal selection of local background is not possible.  We choose to adopt a background
spectrum from a 2\farcm5-radius concentric circle within the ring, where the
intensity appears to be among the lowest values across the FoV.
Given the high intensities along the ring, our
background adoption is not expected to introduce significant bias to the spectral fit.  An absorbed power-law model is found to be an acceptable characterization for almost all the spectra, resulting in absorption column densities of $\sim$$10^{22}{\rm~cm^{-2}}$ and photon-index of $\sim$2, typical of non-thermal emission.  The only exception is the spectrum of the eastern ring (i.e., the shell), for which the power-law model, with a steeper photon-index of $\sim$2.7, is not satisfactory fit.  Fitting the ER spectra with a thermal or thermal+PL model does not lead to a better fit, hence we
present the PL model fit for consistency with other spectra.
The MOS2 spectrum of the plumes northwest to the ring (Fig. 1a;  the
 MOS1 and pn data at this region are contaminated by straylight) can
 also be fitted by an power-law with the index of about 2.2, but
 subject to large uncertainty due to the limited counts. The fit
 results are summarized in Table 1.  Selected spectra and the best-fit
 models are shown in Fig.~3. All the spectra are binned to achieve a signal-to-noise ratio greater than 4 (and a minimum of 30 counts per bin), ranging from about 50 to 350 bins.

We also extract spectra for XMMS J173203-344518 from both the {\xmm}
and {\sl Suzaku} XIS data. The simple pure absorbed power-law model gives a steep photon index, i.e. $\sim$5.1 for the {\xmm} spectra and $\sim$4.7 for the Suzaku spectra.  However, the compact source spectra from both the {\xmm} and Suzaku are better fitted by a combined model (i.e., blackbody plus power-law or two blackbodies). When the {\xmm} and Suzaku data were fitted independently, the temperatures were found to be consistent with being equal ($\sim$ 0.5 keV). To better constrain spectral parameters we jointly fit the {\xmm} and Suzaku spectra. For this joint-fit the column density and temperature were set equal for the {\xmm} and Suzaku spectra (the power-law photon-index is tied for both spectra also); all other parameters were independent. Both the blackbody plus power-law fit ($\chi^2$/d.o.f.=252/260) and the two blackbodies fit ($\chi^2$/d.o.f.=249/260) are good fits. We show the spectra and the blackbody plus powler-law best-fit model in the bottom
panels of Fig. 3. The fit parameters are given in Table 2. 
The compact source shows intrinsic flux in the Suzaku observation a little higher than in the {\xmm} observation. We discuss the nature of this source in \S~4.
  
\subsection{CO spectra}
The three CO spectra (S1: J173200-344200: S2: J173220-345400; S3:
J173250-344600) all show several bright cloud components (Fig. 4). 
Given a near kinematic distance of
$\sim$3.2 kpc for G353.6-0.7, as inferred from the highest HI
absorption feature at a velocity of $\sim$ -20$\pm$4 km s$^{-1}$
(T08), these CO spectra reveal increasing H$_{2}$ column density
($\sim$ 2, 3 and 6 $\times$10$^{21}$ cm$^{-2}$ for S2, S3 and S1,
respectively; see discussion section) towards the Galactic plane. A bright cloud component at -20$\pm$4 km s$^{-1}$ appears in the direction of S1
(i.e. near the intensity peak of the HESS source). This cloud is likely in front
  of the SNR and extends to the nearby bright HII region G353.42-0.37 to produces the deep HI absorption feature at -20$\pm$4 km s$^{-1}$ seen in the HI absorption spectrum from G353.42-0.37. Therefore it is plausible that the extended CO cloud is associated with the remnant and locates at same distance of $\sim$ 3.2 kpc as the HII region.

\section{Discussion and conclusion}
\subsection{Nature of  XMMS J173203-344518}
A number of Galactic TeV objects are suggested to be PWNe \citep*{kar08, gae06}. In view of the proximity between XMMS J173203-344518 and the
projected centers of HESS J1731-347 and SNR G353.6-0.7, it is worth considering a PWN
origin for the $\gamma$-ray emission. 
In \S~3 we described that the spectra of XMMS J173203-344518
can be approximated by a power-law model with a photon index of
4 - 5, which is much steeper than the typical values of PWNe
$\sim$ 1.5 - 2.1 \citep*{li08}. Neither is this index
reminiscent of background active galactic nuclei or Galactic X-ray binaries,
 whose spectra, when fitted with a power-law model, show typical
 photon-index of $\sim$ 1.5 - 2. In addition, XMMS J173203-344518 has no radio counterpart and no extended morphology. The above facts strongly
argue against a possible PWN origin for the HESS source.  

The source's spectra show a harder X-ray tail in the Suzaku observation than in the {\xmm} observation (Fig. 3), so the source is a little brighter in the Suzaku image than in the {\xmm}. We have excluded the possible contamination by the SNR emission, so the flux variability seems be real (the Suzaku observation is 26 days later than the {\xmm}'s). This X-ray variability is possibly consistent with that of a cataclysmic variable (CV) \citep*{pre07}. We have examined the intra-observation (tens of ks in duration) light curves of the source but found no evidence for a flux variation, although we note that CVs do show X-ray variability on longer timescales. XMMS J173203-344518 does not appear in the CVs catalog {\footnote{2006 version:http://heasarc.nasa.gov/W3Browse/all/cvcat.html}}.  

On the other hand, it is not implausible that XMMS J173203-344518 is
the central compact obejct (CCO) associated with SNR
G353.6-0.7. Typically found in young SNRs, CCOs
are characterized by blackbody-like soft X-ray emission with 
temperatures of 0.2-0.5 keV \citep{pav04}. 
It is suggested that CCOs are neutron stars born in supernova explosions with properties different from those of classical rotation-powered pulsars \citep{got05, li07}.  
XMMS J173203-344518 shows a blackbody temperature ($\sim$0.5 keV) higher than most CCOs.  We think that it is possibly a magnetar (i.e. AXPs and SGRs), because magnetars have soft spectra,
a typical blackbody temperature of $\sim$0.5 keV and a typical
luminosity of 10$^{34-36}$ erg s$^{-1}$ \citep{mer08}, also see a magnetar
Catalogue{\footnote{http://www.physics.mcgill.ca/~pulsar/magnetar/main.html}}. Given
a distance of $\sim$ 3.2 kpc, if XMMS J173203-344518 is associated
with the SNR, it will have a  1-10 keV luminosity of
$\sim$ 10$^{34}$ erg s$^{-1}$. This perhaps means it is like AXPs
more than SGRs, because SGRs usually have luminosities 
higher than AXPs, and this source so far  shows no repeated soft $\gamma$-ray emission. 
We examine the {\xmm} and Suzuka light curves of XMMS J173203-344518 
and find no evidence for pulsations at frequencies between 0.05-1 Hz,
a range typical of magnetars.
 Although magnetars often show a hard X-ray (tens of keV) tail that
can be fitted by a shallow power law model, no counterpart of
XMMS J173203-344518 is found in the INTEGRAL/ISGRI map (in the energy
range of 20-40 keV) and INTEGRAL/JEMX map (3-15 keV). In addition, we
notice that there is a little higher column density in the direction of XMMS
J173203-344518 than from other regions of SNR
G353.6-0.7(see Table 1). This may be caused by a local dense
cloud. Future X-ray observations are
needed to test the magnetar scenario for XMMS J173203-344518.   

\subsection{X-ray and $\gamma$-ray emission mechanism}
Based on a statistics of 5 young SNRs, the typical photon index of $\sim$ 2.1 - 3.7 for non-thermal SNRs has been suggested \citep{bam05}. 
It is interesting that the hard X-ray emission of the ring in the HESS
J1731-347/SNR G353.6-0.7 has similar photon index of $\sim$ 2.2. But
the SNR is old due to its extended ($\sim$ 30 arcmin in diameter) and faint radio emission (T08) so it should have a different high-energy mechanism.
Old SNRs have quite different characteristics for the
expected broad-band energy spectrum of accelerated electrons and
protons than young SNRs \citep{yam06}. As an SNR ages, the
acceleration time increases, allowing higher maximum energy of
accelerated particles. However the losses also increase with time. The
net result is that the electron spectrum is loss-limited when the SNR
is older than $\sim10^3$ yr, but the proton spectrum only becomes
loss-limited when the SNR is older than $\sim10^5$ yr. Therefore, in
an old SNR, the $\gamma$- and X-ray emissions are expected to be
dominated respectively by decay of neutral pions and by synchrotron
radiation of secondary electrons from charged pion decay, respectively. These pions
all result from collisions of primary protons and the ISM. One
diagnostic is the ratio of TeV to X-ray synchrotron fluxes.  If the
protons are accelerated at the shock running into a GMC, for example,
the ratio should then be  $\lesssim$10, consistent with the value obtained for HESS J1834-087 \citep{tia07}.  In contrast, if the cloud is only illuminated by the particles, the predicted ratio should then be greater than 100, due to the lack of magnetic field enhancement which would occur in the shock case (hence greatly enhancing the X-ray synchrotron emission).

Based on the best-fit models (Table 1), the 2-10 keV unabsorbed flux
from the ring (including the three knots) is found to be $\approx$ 6.9
$\times$10$^{-12}$ ergs cm$^{-2}$ s$^{-1}$. The $\gamma$-ray flux is
F$_{\gamma(1-10 {\rm TeV})}$ $\approx$ 1.7$\times$
10$^{-11}$ ergs cm$^{-2}$ s$^{-1}$ in the 1-10 TeV band, so the ratio
$R$ = F$_{\gamma}$/F$_{X}$ is $\sim$2.5. A slightly lower value of $R$ is
expected if the X-ray emission from the heavily absorbed western-shell
is also considered.

For an old SNR ($\sim$ 10$^{4}$ yrs for SNR G353.6-0.7), a theoretical model shows that the flux in the 1-10 TeV band is a few times higher than that in 2-10 keV band if the TeV emission is from pion-decays, the X-rays are from synchrotron emissions of secondary electron (in the case of a shocked GMC).
Synchrotron emissions from secondary electrons are about 3 orders of  
magnitude higher than that from primary electrons. Thermal X-ray
emission is not considered in this model. However, there is little
thermal emission originating from an old SNR.

The new Fermi large Area Telescope survey released a strong
$\gamma$-ray source table in the range of 100 MeV - 100 GeV \citep{abd09}. There is no strong GeV source within HESS J1731-347. For a
case that TeV $\gamma$-ray and X-ray emissions originate from a
shocked GMC, the GeV emission flux is about 2 orders of magnitude
lower than that the TeV flux (Fig. 5 of Yamazaki et al.) which is
much low comparing with other cases (Figs. 2-6 of Yamazaki et al.), consistent with the explanation above. 

New CO observations support such an SNR scenario: an extended CO cloud at $\sim$ 20 km s$^{-1}$ is associated with the SNR. From the spectrum of S1, we estimate an approximate H$_{2}$ column density of $\sim$ 4$\times$10$^{21}$ cm$^{-2}$ within the cloud, taking X factor of $\sim$ 2.5 $\times$10$^{20}$ cm$^{-2}$ K$^{-1}$ km$^{-1}$s \citep{sol91}. Assuming the molecular cloud completely covers the SNR (at least over the extended HESS source region), then it has a mass of $M_{H_{2}}$ = $N_{H_{2}}$$\Omega$$d^{2}$(2m$_{H}$/$M_{\odot})$ $\approx$5$\times$$10^{4}$$M_{\odot}$. This is a GMC. 
S1 is located at the peak area of TeV source, and has a higher local H$_{2}$ column density than S2 and S3. This is consistent with the model of a shocked GMC. The evidence of increasing column density from lower-latitude region to higher-latitude region of the SNR revealed by the CO spectra give a plausible reason on the X-rays distribution. The ROSAT X-ray image of SNR G353.6-0.7 in the soft band (0.1-2.4 keV) appears only within the lower-latitude half of the radio remnant where the N$_{H}$ is less than 10$^{22}$ cm$^{-2}$ (e.g. S3 and S2) because absorption blocks soft X-rays but not hard X-rays from the upper half where the N$_{H}$ is larger than 10$^{22}$ cm$^{-2}$.
From the viewing-angle of observers, this gives a physical image that the GMC covers the SNR, so the TeV emissions cover the X-ray and radio emission regions. The coincidence of the X-ray morphology with both the radio and TeV $\gamma$-ray morphologies suggests that they are physically associated.

\begin{acknowledgements}
%WWT thanks the Natural Science Foundation of China for support. 
WWT and DL acknowledge support from the Natural Sciences and Engineering Research Council of Canada. We thank Drs. A. Bamba, C. Heinke and S. Zhang for their help on analyzing Suzaku and INTEGRAL data. RY was supported in part by a Grant-in-aid from the Ministry of Education, Culture, Sports, Science, and Technology of Japan (No. 21740184). Upon the submission
of our manuscript, we became aware of the recent work by \citet{ace09} on the HESS J1731-347/SNR G353.6-0.7 system with independent approaches.
\end{acknowledgements}

\clearpage 

\begin{deluxetable}{cccccc}
\tabletypesize{\footnotesize}
\tablecaption{Spectral fit results from {\xmm} image}
\tablewidth{0pt}
\tablehead{
\colhead{Parameter} &
\colhead{K1} &
\colhead{K2} &
\colhead{K3} &
\colhead{ER} &
\colhead{WR}\\ 
}
\startdata
$\chi^2/d.o.f.$ \dotfill &  29/22 & 147/109  & 79/77 & 88/54  & 155/165\\ 
$N_{\rm H}$ ($10^{22}{\rm~cm^{-2}}$) \dotfill  & 1.44$^{+0.55}_{-0.26}$ &1.21$^{+0.15}_{-0.12}$ & 0.86$^{+0.11}_{-0.11}$ &1.08$^{+0.19}_{-0.17}$ & 1.85$^{+0.20}_{-0.16}$\\
Photon index \dotfill & 1.98$^{+0.44}_{-0.22}$& 2.15$^{+0.14}_{-0.13}$ & 2.25$^{+0.15}_{-0.15}$ & 2.72$^{+0.29}_{-0.26}$ &2.39$^{+0.16}_{-0.13}$ \\
Norm (PL; $10^{-4}$) \dotfill & 1.1$^{+0.9}_{-0.3}$ & 5.3$^{+1.1}_{-0.8}$ & 4.3$^{+0.9}_{-0.7}$  &12$^{+5}_{-3}$ & 25$^{+6}_{-4}$ \\
Flux$_{2-10keV}$($10^{-13}{\rm~ergs~s^{-1}~cm^{-2}}$) \dotfill &  2.8  & 11.0 & 7.8 & 11.5 & 36.3  \\ 
\enddata
\tablecomments{Regions of spectral interest: K1, K2 and K3 -three bright knots; ER - the eastern half of the ring excluding K2 and K3; WR - the western half
  of the ring excluding K1;  See details in text. Quoted uncertainties are at 90$\%$ confidence level. An absorbed power-law model is used to fit all spectra.}
\label{tab:spec}
\end{deluxetable}

\begin{deluxetable}{ccccccc}
\tabletypesize{\tiny}%footnotesize}
\tablecaption{Spectral fit results from the X-ray compact source}
\tablewidth{0pt}
\tablehead{
\colhead{Parameter} &
\colhead{BB$^a$} &
\colhead{PL$^a$} &
\colhead{BB+PL$^a$} &
\colhead{BB+BB$^a$} &
\colhead{BB+PL$^b$} &
\colhead{BB+BB$^b$}\\
}
\startdata
$\chi^2/d.o.f.$ \dotfill & 120/78 & 118/78 & 76/78 & 82/78 & 252/260 & 249/260 \\
$N_{\rm H}$ ($10^{22}{\rm~cm^{-2}}$) \dotfill & 1.36$\pm$0.06& 3.20$\pm$9.09 & 2.56$\pm$0.24 &1.53$\pm$0.08 & 2.80$^{+0.26}_{-0.26}$ &1.96$^{+0.27}_{-0.22}$ \\
Photon index \dotfill &  & 4.69$\pm$0.08 &4.40$\pm$0.32& & 4.84$^{+0.48}_{-0.38}$ &  \\
Norm (PL;$10^{-2}$) \dotfill &  &5.27$\pm$0.56 &1.91$\pm$0.91 &  & $^c$2.41$^{+1.41}_{-0.93}$/$^d$2.38$^{+1.50}_{-0.96}$  &  \\
Temp$_1$ (keV) \dotfill & 0.52$\pm$0.01  & &0.48$\pm$0.03&0.47$\pm$0.02&0.50$^{+0.02}_{-0.02}$&0.31$^{+0.07}_{-0.05}$\\
Norm$_1$ (BB) \dotfill & 9.51$\pm$0.87 & &9.21$\pm$3.35&14.92$\pm$2.51&$^c$6.0$^{+2.0}_{-1.6}$/$^d$9.1$^{+2.9}_{-2.3}$&$^c$73.9$^{+171}_{-46.2}$/$^d$71.5$^{+165}_{-43.5}$\\
Temp$_2$ (keV) \dotfill &&&&2.37$\pm$-0.49&&0.57$^{+0.06}_{-0.03}$\\
Norm$_2$(BB;$10^{-2}$) \dotfill &&&&1.37$\pm$1.66&&$^c$3.6$^{+2.3}_{-2.2}$/$^d$4.8$^{+2.8}_{-2.8}$\\
Flux \dotfill & 62  & 313 & 168 & 71& 168$^e$/184$^f$ & 76$^e$/87$^f$ \\
\enddata
\tablecomments{The fitting models: BB - Blackbody; PL - power-law. The spectral extraction area has a radius of $\sim$ 3 arcmin in Suzaku image larger than that (40 arcsec) in {\xmm} image; Quoted uncertainties are at 90$\%$ confidence level. $^a$ Spectra obtained from the Suzaku. $^b$ Spectra obtained from the joint {\xmm} and Suzaku. $^c$ fitting parameters to the {\xmm}. $^d$ fitting parameters to the Suzaku. $^e$ 1-10keV ($10^{-13}{\rm~ergs~s^{-1}~cm^{-2}}$) intrinsic flux from the {\xmm}. $^f$ 1-10keV ($10^{-13}{\rm~ergs~s^{-1}~cm^{-2}}$) intrinsic flux from the Suzaku. }
\label{tab:spec}
\end{deluxetable}

\begin{figure}
\vspace{40mm} \begin{picture}(80,80)
\put(310,-10){\includegraphics{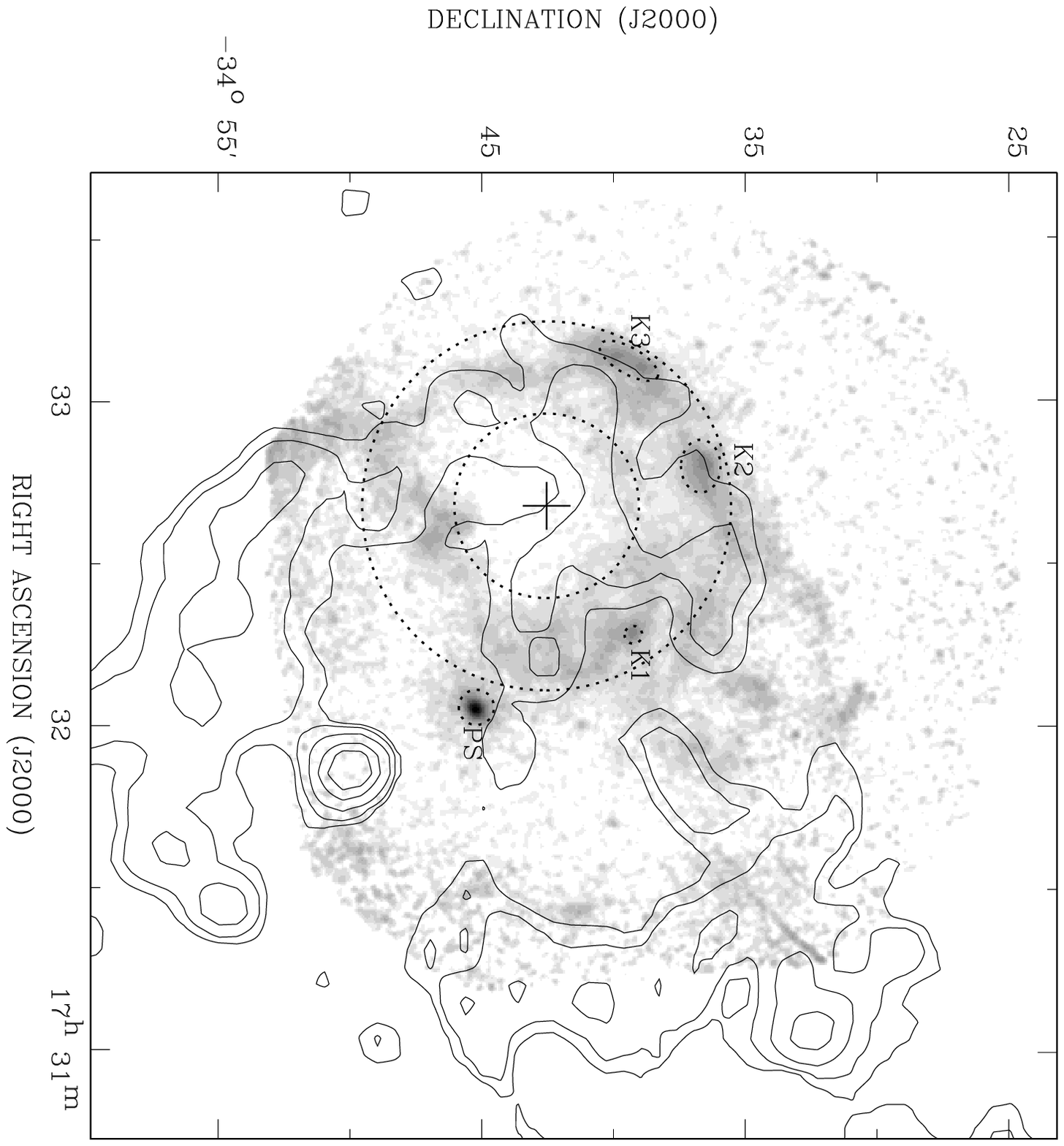}}
\put(580,-10){\includegraphics{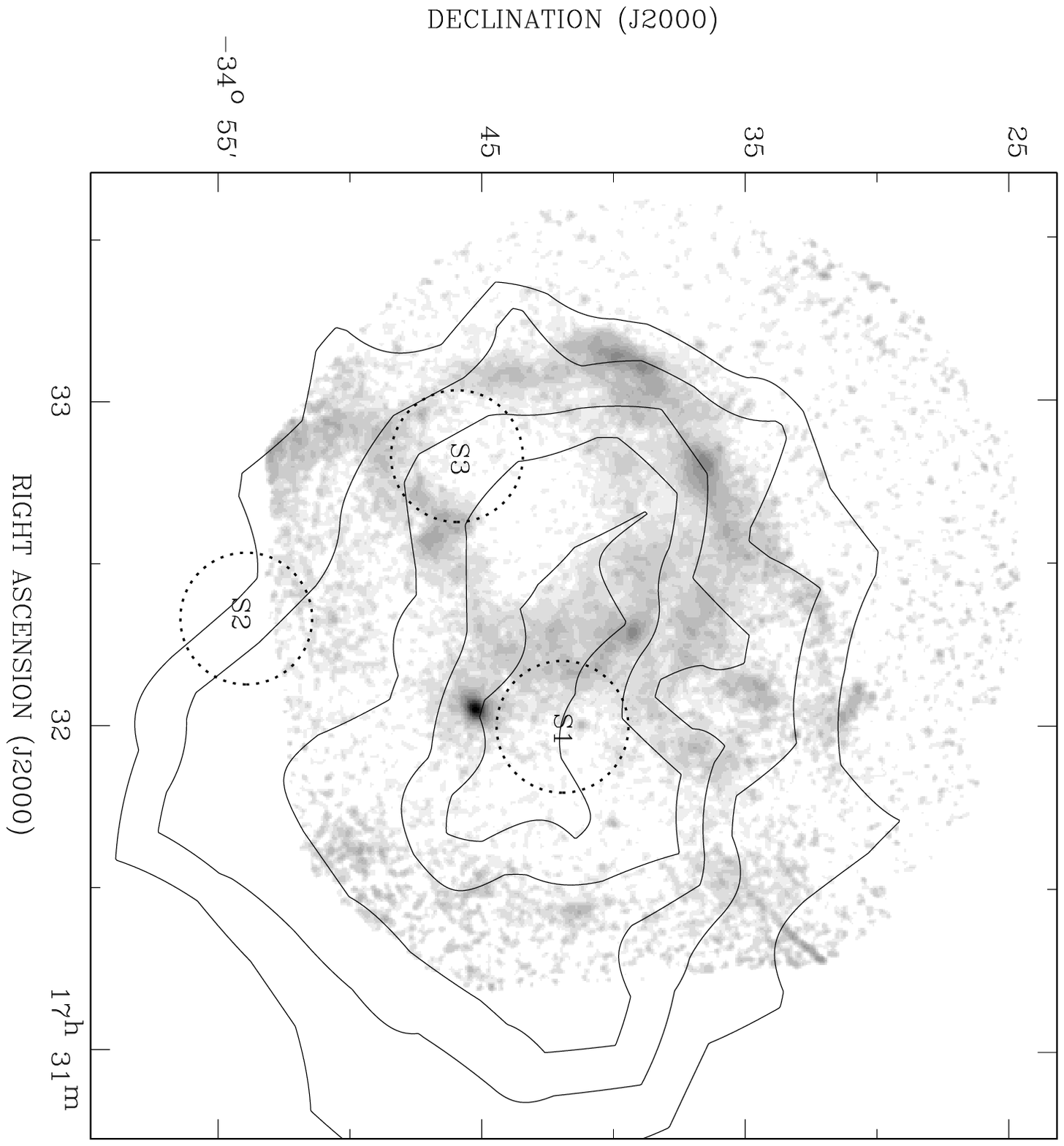}} 
\end{picture}
\caption[xx]{Smoothed {\xmm} 0.8-7 keV intensity (greyscale) images overlaid with contours
  of the VLA radio continuum emission (a) and HESS $\gamma$ray emission
  (b), respectively. The greyscale is logarithmically coded between (0.5-250)$\times10^{-4}{\rm~cts~s^{-1}~arcmin^{-2}}$. a: The region enclosed by the two dotted circles outlines the ring-like feature, while the small ellipse /circle
  outline the bright knots (K1, K2 and K3) and the compact source (CS). b: The circles show the regions where the CO spectra are extracted from.}
\end{figure} 

%\clearpage 

\begin{figure}
\vspace{60mm} \begin{picture}(80,80)
\put(320,-30){\includegraphics{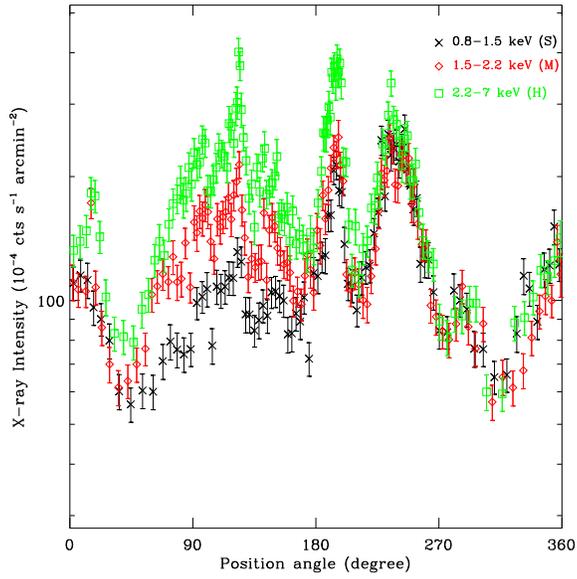}}
 \end{picture}
\caption[xx]{Azimuthal X-ray intensity distributions of the ring,
  extracted from an annulus, with inner-to-outer radii of
  3\farcm5-7$^\prime$, centering at a position marked by a cross in
  Fig.~1a. Position angles are measured counterclockwise from south.}
\end{figure} 

\clearpage 

\begin{figure}
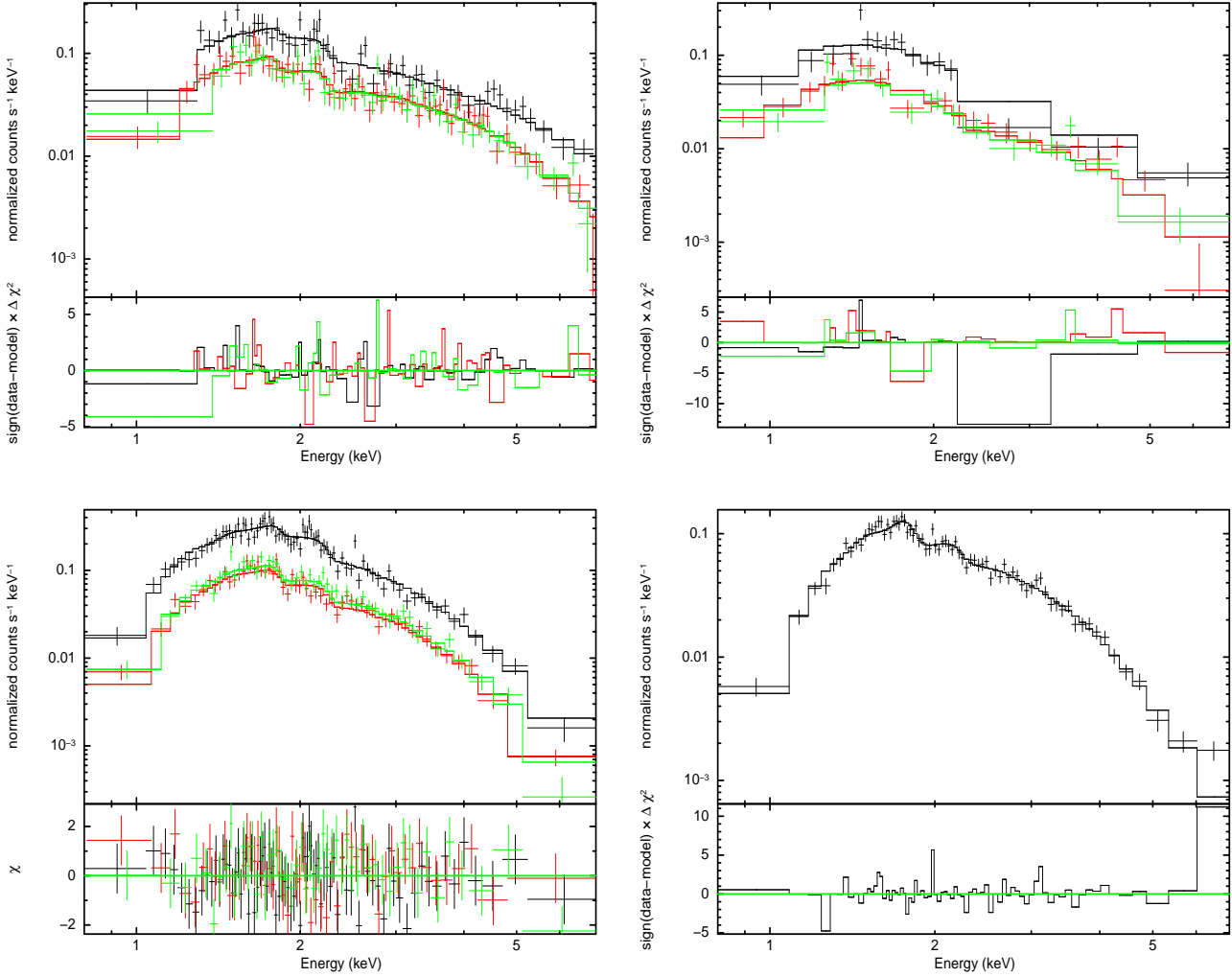

\vspace{120mm} 
\begin{picture}(80,80)
\put(-30,450){\includegraphics{f3a.ps}}
\put(220,450){\includegraphics{f3b.ps}}
\put(-30,250){\includegraphics{f3c.ps}}
\put(220,250){\includegraphics{f3d.ps}}
\end{picture}
\caption[xx]{Spectra ({\sl black}: EPIC-pn or Suzaku XIS (right panel of bottom row); {\sl red}: EPIC-MOS1; {\sl green}: EPIC-MOS2) extracted from the western ring (left panel of the first row), the eastern ring (right panel of the first row) and the compact source (bottom row) and the best-fit models
  (see text). The {\sl XMM-Newton} spectra are binned to achieve a signal-to-noise
  ratio greater than 4 and a minimum of 30 counts per bin, while the
  {\sl Suzaku} spectrum is binned with a minimum of 100 counts.}
\end{figure}

\clearpage

\begin{figure}
\vspace{120mm}
\begin{picture}(80,80)
\put(-20,400){\includegraphics{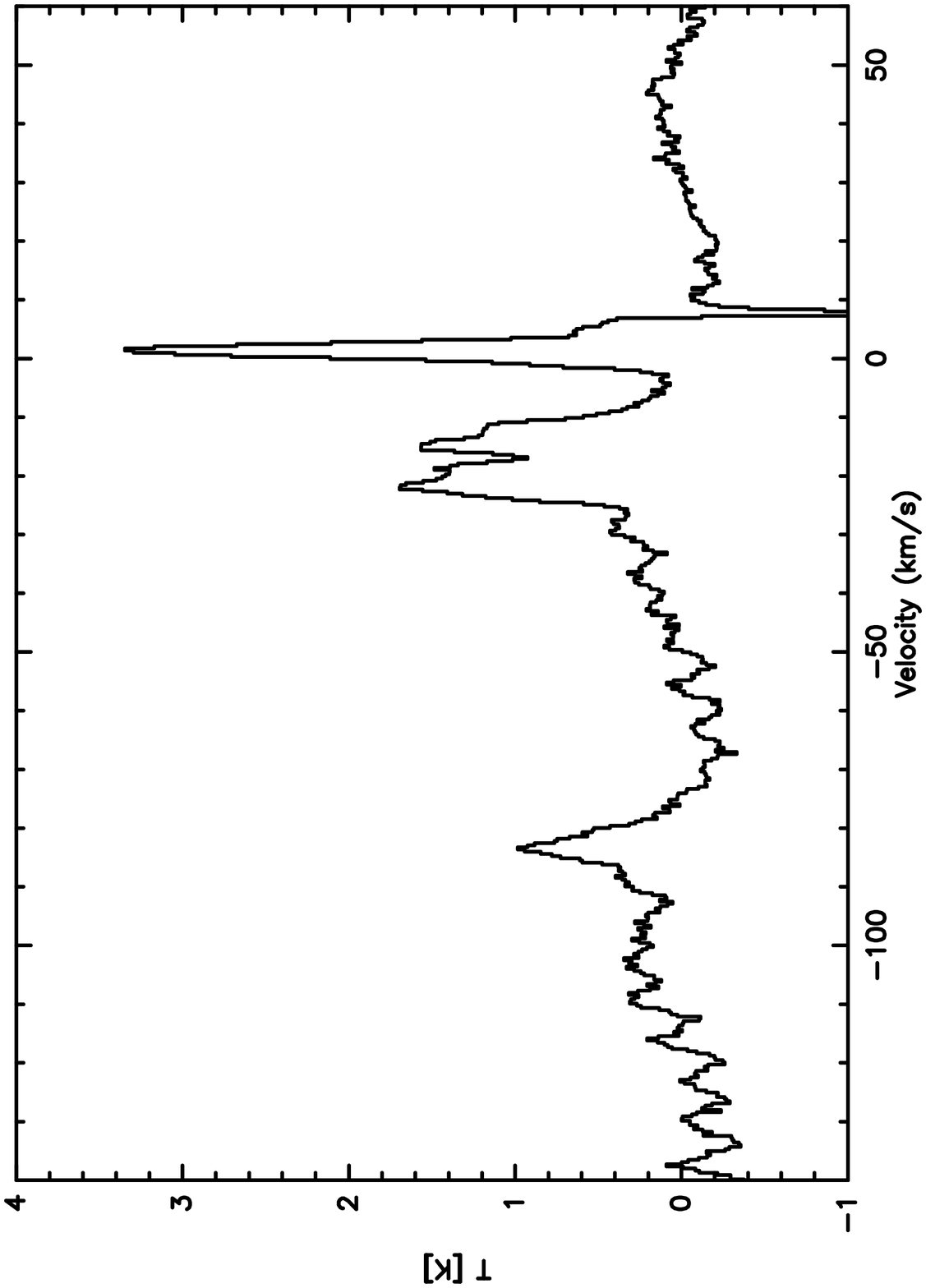}}
\put(240,400){\includegraphics{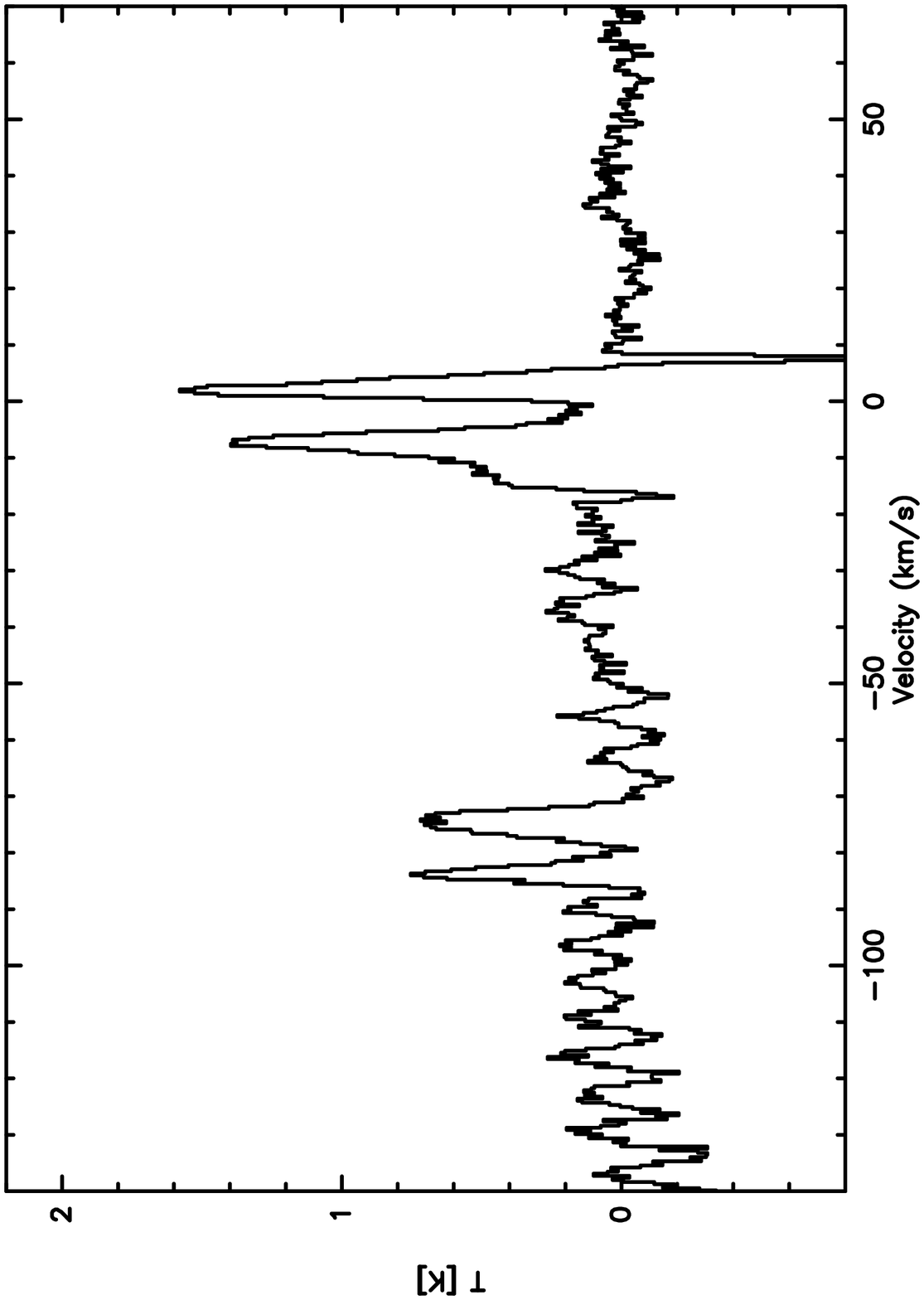}}
\put(-20,200){\includegraphics{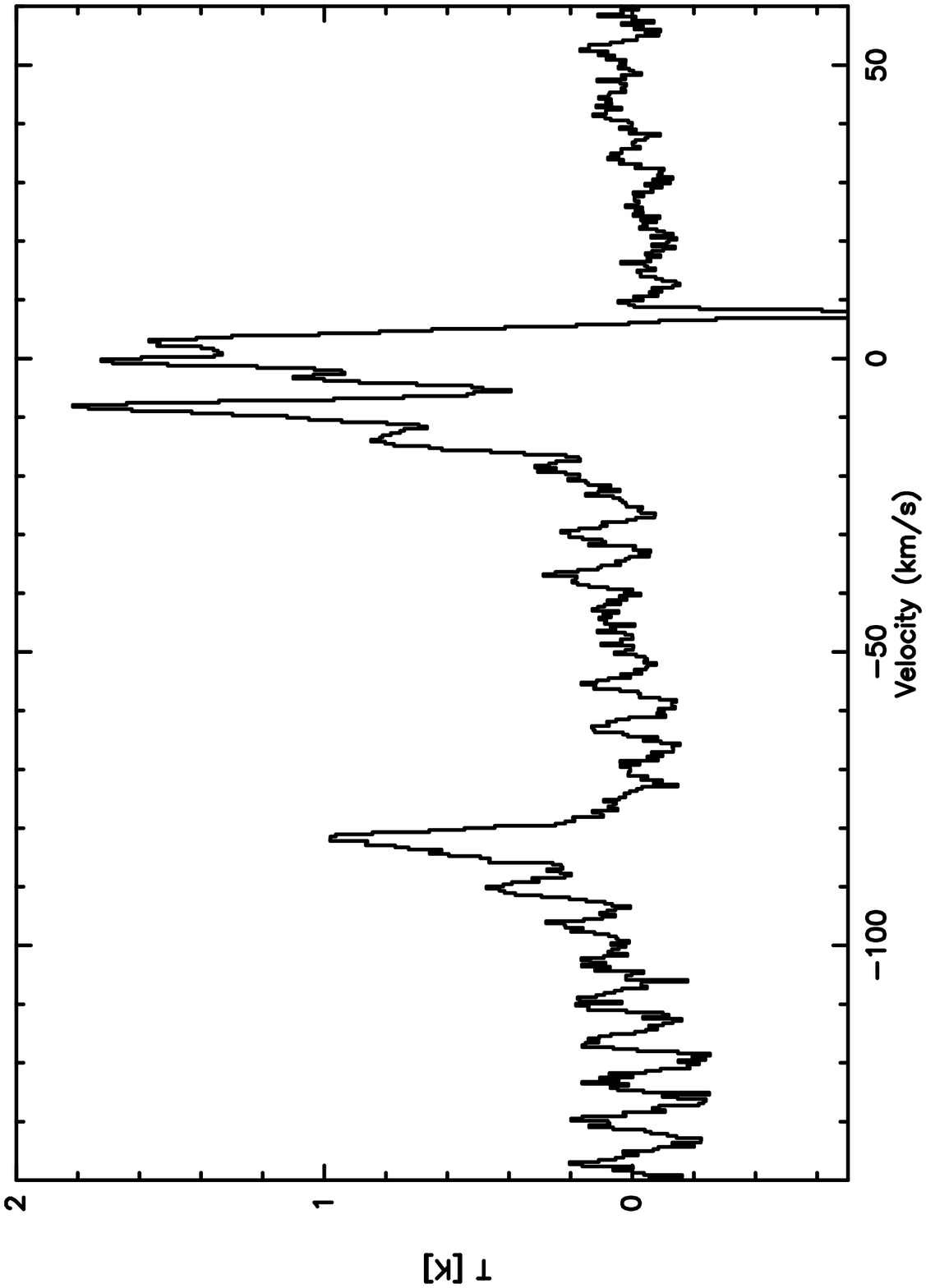}}
\end{picture}
\caption[xx]{The CO spectra from the TeV source peak (S1: the left panel of the first row) and the X-ray peak sites (S3: right of the first row; S2: the bottom row)}
\end{figure}

\end{document}